\documentclass{aastex63}

\date{\today}
\submitjournal{ApJ}
\usepackage{natbib}
\usepackage{longtable}
\usepackage{amssymb}
\usepackage{rotating}
\usepackage{hyperref}
\usepackage[switch,columnwise]{lineno}

\newcommand{\msun}{M_\odot}

\newcommand{\kms}{{\rm km~s^{-1}}}
\newcommand{\jj}{{\rm cm^2~s^{-1}}}

\submitjournal{ApJ}

\shorttitle{Super-Fast Rotating Jets}
\shortauthors{Matsushita et al.}
\begin{document}

%----- Title
\title{Super-Fast Rotation in the OMC 2/FIR 6b Jet}% 32/90 char
%\title{Super-rotating Protostellar Jets}% 32/90 char
%----- Author
\correspondingauthor{Yuko~Matsushita}
\email{mathcat.e2.718@gmail.com}

\author[0000-0002-9091-963X]{Yuko~Matsushita}
\affiliation{National Astronomical Observatory of Japan, 2-21-1 Osawa, Mitaka, Tokyo 181-8588, Japan}
\affiliation{Department of Earth and Planetary Sciences, Faculty of Sciences, Kyushu University, Fukuoka 819-0395, Japan}

\author[0000-0002-7287-4343]{Satoko~Takahashi}
\affiliation{NAOJ Chile Observatory, Alonso de C{\'{o}}rdova 3788, Oficina 61B, Vitacura, Santiago, Chile}
\affiliation{Joint ALMA Observatory, Alonso de C{\'{o}}rdova 3107, Vitacura, Santiago, Chile}
\affiliation{Department of Astronomical Science, School of Physical Sciences, The Graduate University for Advanced Studies (SOKENDAI), 2-21-1 Osawa, Mitaka, Tokyo 181-8588, Japan}

\author{Shun~Ishii}
\affiliation{National Astronomical Observatory of Japan, 2-21-1 Osawa, Mitaka, Tokyo 181-8588, Japan}
\affiliation{Department of Astronomical Science, School of Physical Sciences, The Graduate University for Advanced Studies (SOKENDAI), 2-21-1 Osawa, Mitaka, Tokyo 181-8588, Japan}

\author{Kohji~Tomisaka}
\affiliation{National Astronomical Observatory of Japan, 2-21-1 Osawa, Mitaka, Tokyo 181-8588, Japan}
\affiliation{Department of Astronomical Science, School of Physical Sciences, The Graduate University for Advanced Studies (SOKENDAI), 2-21-1 Osawa, Mitaka, Tokyo 181-8588, Japan}

\author{Paul T. P. Ho}
\affiliation{Academia Sinica Institute of Astronomy and Astrophysics, P.O. Box 23-141, Taipei 106, Taiwan}
\affiliation{East Asian Observatory, Hilo 96720, HI, USA}

\author[0000-0003-2251-0602]{John M. Carpenter}
\affiliation{Joint ALMA Observatory, Alonso de C{\'{o}}rdova 3107, Vitacura, Santiago, Chile}

\author{Masahiro N.~Machida}
\affiliation{Department of Earth and Planetary Sciences, Faculty of Sciences, Kyushu University, Fukuoka 819-0395, Japan}

%\author{{Yuko~Matsushita}$^{1,2}$, {Satoko~Takahashi}$^{3,4,5}$, {Shun~Ishii}$^{1,5}$,  {Kohji~Tomisaka}$^{1,5}$, {Paul T. P. Ho}$^{6,7}$, {John Carpenter}$^{4}$, \& {Masahiro N.~Machida}$^{2}$}
%\affil{$^{1}$ National Astronomical Observatory of Japan, 2-21-1 Osawa, Mitaka, Tokyo 181-8588, Japan\\
%$^{2}$ Department of Earth and Planetary Sciences, Faculty of Sciences, Kyushu University, Fukuoka 819-0395, Japan\\
%$^{3}$ NAOJ Chile Observatory, Alonso de C{\'{o}}rdova 3788, Oficina 61B, Vitacura, Santiago, Chile\\
%$^{4}$ Joint ALMA Observatory, Alonso de C{\'{o}}rdova 3107, Vitacura, Santiago, Chile\\
%$^{5}$ Department of Astronomical Science, School of Physical Sciences, The Graduate University for Advanced Studies (SOKENDAI), 2-21-1 Osawa, Mitaka, Tokyo 181-8588, Japan\\
%$^{6}$ Academia Sinica Institute of Astronomy and Astrophysics, P.O. Box 23-141, Taipei 106, Taiwan\\
%$^{7}$ East Asian Observatory, Hilo 96720, HI, USA
%}

\begin{abstract}
We present ALMA CO ($J$=2--1) and 1.3 mm continuum observations of the high-velocity jet associated with the FIR 6b protostar located in the Orion Molecular Cloud-2. 
We detect a velocity gradient along the short axis of the jet in both the red- and blue-shifted components.
% The velocity gradient is highly systematic in the red-shifted component, in which the velocity changes along the short axis of the jet. 
The position-velocity diagrams along the short axis of the red-shifted jet show a typical characteristic of a rotating cylinder. 
We attribute the velocity gradient in the red-shifted component to rotation of the jet. 
The rotation velocity ($>20\,\kms$) and specific angular momentum ($>10^{22}\,\jj$) of the jet around FIR 6b are the largest among all  jets  in which  rotation has been observed.
By combining disk wind theory with our observations, the jet launching radius is estimated to be in the range of $2.18-2.96$\,au. 
% JMC: I think this is too many significant figures given the uncertainties.
The rapid rotation, large specific angular momentum, and a launching radius far from the central protostar can  be explained by a magnetohydrodynamic disk wind that contributes to the angular momentum transfer in the late stages of protostellar accretion. 
% JMC: I suggest  to get rid of the word "only" in the above sentence. Make sure that you are happy with the last sentence in general since I made other changes as well.
\end{abstract}


\section{Introduction} 
\label{sec:intro}
Protostellar jets are  striking phenomena in star-forming regions and are considered to be an essential ingredient in the star formation process. 
Stars form in gravitationally collapsing clouds. 
The mass of a protostar at its birth is approximately a Jovian mass, or 0.1\% the mass of the sun \citep{larson69,masunaga00}.  
After its birth, a protostar acquires additional mass by accreting material from a surrounding disk embedded in an infalling envelope.
% that is a remnant of the natal cloud of the star.
However, the rotation (or large angular momentum) of the disk prevents mass accretion onto the protostar from the circumstellar disk.
Protostellar jets are believed to expel the excess angular momentum from the circumstellar region and allow accretion onto the protostar.

Observations of rotation in protostellar jets can be used to estimate the jet driving region assuming the Keplerian rotation of the circumstellar disk. 
In addition,  we can identify the jet driving mechanism using the law of conservation of  angular momentum \citep[e.g.][]{mestel68}. 
However, very high spatial resolution observations are necessary to measure jet rotation because protostellar jets are well collimated. 
Thus, jet rotation has been observed in only a limited number of cases. 
Since the first discovery of  jet rotation \citep{bacciotti02}, several studies have observed rotation in protostellar jets \citep{Launhardt09, bjerkeli16,chen16,tabone17,lee17,zhang18}. 
Recently, \citet{hirota17} clearly showed that rotation is present in the protostellar outflow driven by Orion Source I and estimated an outflow launching radius of $>$ 10~au from the central protostar. 
% JMC: Please verify the above sentence is correct. I made it more specific by adding > 10 au.
Their study suggests that protostellar outflows are directly driven from the intermediate region of the circumstellar disk by  magnetohydrodynamic  \citep[see also][]{hirota20}.  

In this study, we report the detection of fast rotating jets driven from the protostar FIR 6b (HOPS 60) in the Orion Molecular Cloud-2 \citep[OMC-2;][]{mezger90,chini97,furlan16}.
The jet rotation velocity exceeds $20\,\kms$ and the specific angular momentum of the jet is as large as $\sim10^{22}\,\jj$, which hitherto is the largest value that has been observed in protostellar jets.
The extraordinary large rotation velocity and specific angular momentum can be explained by a magnetohydrodynamic disk wind \citep{blandford82}.
This is clear evidence that magnetic fields play a crucial role for protostellar evolution and that angular momentum is removed by protostellar jets.

% The target FIR 6b (HOPS 60) is located in Orion Molecular Cloud-2 (hereafter we call this object FIR 6b) and  
%  JMC: I did not think the above was necessary.
FIR 6b is a Class 0 intermediate mass protostar \citep{chini97, furlan16, takahashi08}. 
The distance to FIR 6b is $d\sim$392 pc \citep{tobin20} and the systemic velocity is $v_{\rm sys} \sim11~\kms$ in the Local Standard of Rest (LSR) \citep{ikeda07}.
FIR 6b has a bolometric luminosity of $L_{\rm bol}$ = 21.9  $L_\odot$ \citep{furlan16}. 
%%$L_\odot$ and a total luminosity of $L_{\rm tot}$ = 78.7
A low-velocity ($\sim5\,\kms$) bipolar outflow from FIR 6b extends from the northeast (blueshifted lobe) to the southwest (redshifted lobe) \citep{takahashi08,shimajiri09}. %%with a velocity of  (7.7 to 12.6 $\kms$).
% JMC: Verify that the above sentence is correct
A small amount of gas remains in the envelope around  FIR 6b \citep{chini97, furlan16, shimajiri09}.
In addition, a disk-like structure with a size of $\sim126$\,au and $\sim56$\,au were also inferred from continuum observations at 0.87 mm by  the Atacama Large Millimeter/submillimetre Array (ALMA) and 9 mm by the Very Large Array (VLA), respectively \citep{tobin20}.
%\textcolor{red}{In addition, the disk-like structure with a size of $\sim50$\,au was also inferred from the shape of the spectral energy distribution\citep{furlan16}. }
Thus, FIR 6b is expected to be in the late main accretion phase and considered to be a Class 0 protostar.
%  JMC: I did not think this last sentence was needed since the first sentence of the paragraph says it is a Class 0 protostar.

The structure of this paper is as follows.
We describe the observations in \S2.
The observational results are presented in \S3. 
We investigate the jet rotation and the driving mechanism in \S4.    
The parameters and properties of the jets are described in \S5. 
We discuss the effect of the inclination angle and the low-velocity component in \S6.
A summary is given in \S7. 

\section{Observation} \label{sec:obs}
Mosaicking observations of the OMC-2/FIR 6 region in CO ($J$=2--1; 230.538 GHz) and the 1.3 mm continuum were obtained with the ALMA 12-m array on 19 April 2018 and with the ACA 7-m array (Morita array) on 7, 10, 11, and 17 January 2018. 
The data were obtained through the Cycle 5 program 2017.1.01353.S (PI: S. Takahashi).

The mosaics cover a $3.8' \times 3.9'$ area centered at (R.A., Dec.) = ($05^{h}35^{m}21^{s}.7$, $-05^{\circ}12'51\farcs0$) with Nyquist sampling.
%We focus on FIR 6b sources in the OMC-2/FIR 6 region and the 1.3 mm continuum peak position is R.A. = $05^{h}35^{m}23^{s}.343$, Dec = $-05^{\circ}12'03''.970$. 
In the  12-m array and  7-m array observations, the mosaic consists of 108 and 42 pointings with an on-source integration time per pointing of 20 and 260 seconds, respectively.
% JMC: I modified the above sentence. Make sure it is correct.
The overview of the full survey at OMC-2/FIR 6 region will be presented in a separate paper (Matsushita et al. in prep 2021).
In this paper, we present the results in the area 
%(R.A., Dec.) = ($05^{h}35^{m}23^{s}.34$, $-05^{\circ}12'03''.970$) 
centered on OMC-2/FIR 6b.

The correlator was configured to have 4 spectral windows, each  observed in dual polarization mode. Two spectral windows were centered at
233.100 and 215.200 GHz with a bandwidth of 1875 MHz each to measure the continuum emission.
% Two spectral windows with a 1875 MHz width are allocated to the continuum observations centered at 233.100 and 215.200 GHz.
The other two spectral windows were centered on CO ($J$=2--1) and SiO ($J$=5--4) with a bandwidth of 938 MHz  and a spectral resolution of 244 kHz (0.64  $\rm km\ s^{-1}$ for CO $J$ = 2--1).
%  JMC: 244 kHz corresponds to 0.32 km/s, not 0.64 kms. Is 244 kHz really the resolution or the channel spacing? If you did not bin the data, then the velocity resolution is about twice the channel spacing.
The SiO emission is not detected in FIR 6b.
The arrays consisted of 44 and 11 antennas for the  12-m array and the  7-m array observations, respectively, with projected baseline coverage from 15.1 to 500.2 m and from 8.9 to 48.9 m.
The full-width-at-half-maximum (FWHM) primary beam size is 25\farcs2 and 43\farcs2, and the system temperatures ranged between 70 to 180 K and 60 to 210 K for the  12-m array and the  7-m array, respectively.
The flux, gain, and bandpass calibrators were J0423-0120, J0541-0211, and J0423-0120 for the  12-m array, and J0522-3627 and J0607-0834, J0542-0913, and J0423-0120 for the  7-m array.

The data were calibrated using the Common Astronomy Software Application \citep[CASA;][]{mcm07}, version 5.1.1 with the ALMA pipeline and imaged with CASA version 5.4.0.
The CO line emission was separated from the continuum emission using “uvcontsub” task in CASA.
The  line--free regions in the CO and SiO spectral windows were combined with the two continuum windows to provide a total effective continuum bandwidth of 5.75 GHz.
% JMC: Make sure the above sentence is correct.

The CO data from the 7-m array and the 12-m array were combined using the CASA task “concat” with a weight of 20:1. 
% JMC: Is the weight correct? Was the 70m array data weighted 20 times more than the 12-m array data?
Only the ALMA 12m array data were used to image of the 1.3 mm continuum emission.
The CO and 1.3 mm continuum data were imaged by the CASA task “tclean” with a robust weight of 0.5.
The primary beam correction was applied.
The synthesized beam size of the CO cube has a FWHM size of $4\farcs5 \times 2\farcs5$  with a position angle of $-87.0^{\circ}$, and the continuum image has a beam size of $1\farcs0 \times 0\farcs78$ with the position angle of $-73.6^{\circ}$,
% of the final cube of the CO data and image of continuum data is $4\farcs5 \times 2\farcs5$  with the position angle of $-87.0^{\circ}$ and $1.0 \times 0.78$ arcsec with the position angle of $-73.6^{\circ}$, respectively.
The achieved noise levels are 30 $\rm{mJy\ beam^{-1}}$ for the CO image cube with a velocity resolution of 2.5 $\rm km\, s^{-1}$ and 0.5 $\rm{mJy\ beam^{-1}}$ for the 1.3 mm continuum image.

\section{Results} \label{sec:results}
\begin{figure}[!ht]
\centerline{\includegraphics[width=0.9\textwidth]{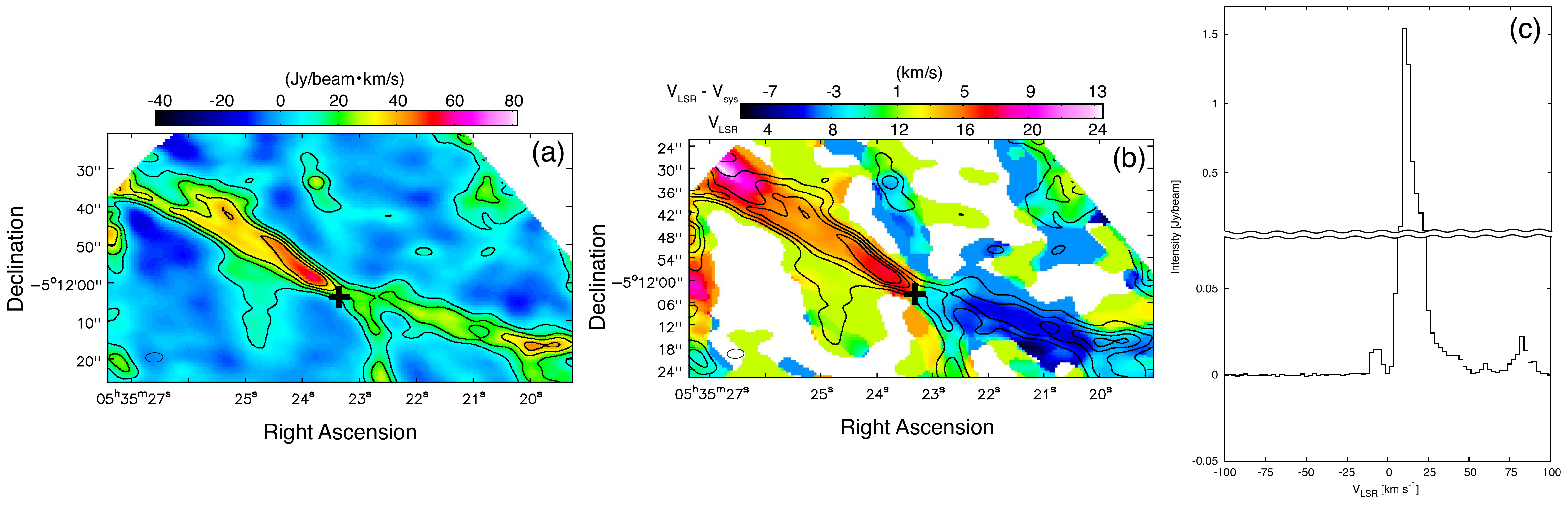}}
%%\centerline{\includegraphics[width=0.9\textwidth]{f9_pb.eps}}
\caption{ 
(a) CO ($J$=2--1) integrated intensity map of the low velocity component in the FIR 6b jet.
The emission has been integrated between velocities of 0 and $30\,\kms$, but excluding  emission in the systemic velocities between 7.5 and $12.5\,\kms$.
The plus symbol at the center corresponds to the position of FIR 6b.
The contour levels are at 5$\sigma$, 10$\sigma$, 15$\sigma$, and 20$\sigma$ (1 $\sigma$ $=$ 2.0 $\rm{Jy\ beam^{-1}} \cdot \rm{km\ s^{-1}}$).
The black open ellipse in the bottom left corner indicates the synthesized beam.
(b) Same as in panel a, but for  the mean CO ($J$=2--1) velocity.
The contours represent the integrated intensity shown in panel a.
(c) The CO ($J$=2--1) line profile.
% JMC: Panel (c) needs more information. Is this the average and integrated line profile over the area shown, or toward a particular line of site.?
}
\label{f1}
\end{figure}

%%\subsection{Low- and High-velocity Components}
% We observed the region around FIR 6b using ALMA.
Figures~\ref{f1}{\it a} and \ref{f1}{\it b} show the integrated intensity and mean velocity maps  of the low velocity components ($5-20\,\kms$ with respect to the systemic velocity) toward FIR 6b. 
Cavity-like structures are seen in both ends of the red-shifted and blue-shifted jets in Figure~\ref{f1}{\it a}.
% JMC: I changed "heads" to "ends"
The moment 1 map (Fig.~\ref{f1}{\it b}) shows both the red-shifted (northeast) and blue-shifted (southwest) components identified previously \citep{takahashi08,shimajiri09}.
Blue-shifted emission is detected between velocities of $-20$ and 10\,$\kms$, while red-shifted emission is detected between 10 and 100\,$\kms$ (see Figure~\ref{f1}{\it c}).
Figure~\ref{f1}{\it c}  also shows that both the red- and blue-shifted components contain two local emission peaks corresponding to the high velocity components. 
% JMC: Make sure I modified this sentence ccorrectly. I did not understand what this sentence means though the figure shows only the emission from the low velocity component, so how can be peaks be due to the high velocity  component? Or are you referring to Figure 1c? But I do not see two peaks on the blue shifted side.

For the first time, we could confirm the high-velocity components (or high-velocity jets) driven from FIR 6b, which is distributed from the northeast to the southwest direction centered around FIR 6b as in the low-velocity components (Fig.~\ref{f1}).  
Figure~\ref{f2} shows the image of the high velocity CO ($J$=2--1) emission, in which the high velocity components are integrated between 32.5 to 97.5$\,\kms$ for the red-shifted jet, and between 17.5 to 0$\,\kms$ for the blue-shifted jet. The red- and blue-shifted components of the high velocity emission are distributed in the same general direction as the low-velocity components (see Figure~\ref{f1}).
The integrated high-velocity CO emission shows a well-collimated structure, within which several knots corresponding to local emission peaks indicated by arrows are embedded in the high-velocity jets.

The length of the jet, measured from the end of the blue--shifted component to the end of the red--shifted component, is $\sim48,000$\,au. 
% JMC: Make sure above sentence is correct.
The jet width is $\sim2400-5000$\,au depending on the distance from the central protostar.  
Although the CO emission in the southwest direction is slightly distorted near FIR 6b, the two components of the jet have a roughly  symmetric structure around FIR 6b (Fig.~\ref{f2}). 
The asymmetrical velocity structure in the upper- and lower-side of jets is sometimes seen in other observations \citep{matsushita19}.
%  JMC: I am not sure what asymmetrical velocity structure you are referring to.
Recent simulations indicate that an asymmetrical jet (velocity) can be realized by the temporal asymmetric mass accretion from the infalling envelope onto the circumstellar disk \citep{matsumoto17}. 
The northeast side of the jet moves away from us with a velocity range of $22.5~\kms$ to $87.5~\kms$ with respect to the systemic velocity, while the southwest side is coming toward us with a velocity of $10~\kms$ to $27.5~\kms$ with respect to the systemic velocity.
% JMC: Is this what you mean by asymmetric velocites? If so, this sentence should be moved earlier.
Thus, the northeast part corresponds to the red-shifted component (or the red-shifted jet), while the southwest part is the blue-shifted component (or the blue-shifted jet). 
The propagation direction of the high-velocity components is the same as that of the low-velocity components reported in previous studies \citep[][see also Fig.~\ref{f1}]{takahashi08, shimajiri09}.
% JMC: The above two sentences repeat information in the earlier paragraphs and could be removed.
Since the jets have a velocity range of $10~\kms$ to $87.5~\kms$ with respect to the systemic velocity
% in the redshifted components
% JMC: Note that I deleted "in the redshifted components" since the velocity range of 10-87.5 km/s refers to both the red and blue components. The red component on its own is 22.5 to 87.5 km/s.
(Fig.~\ref{f1}{\it c}), the intrinsic jet peak velocity might exceed $100~\kms$ after correcting for jet inclination angle (for details, see \S\ref{sec:parameters} and \S\ref{sec:inclination}).
%%  of $i=80^{\circ}$ with respect to the line of sight.
%%Since the jets have a velocity range of \textcolor{red}{$10-80~\kms$ with respect to the systemic velocity} in both blue- and red-shifted components, \textcolor{red}{the line of sight jet velocity width} exceeds $100~\kms$. 
However, the blue-shifted jet has slower velocity than the red-shifted one, and no high velocity component has been observed in the southeast direction.
% JMC: I did not understand what is meant by "no high velocity component has been observed in the southeast direction.". Figure 2 shows there is a high velocity componenet. Do you mean there is not as extreme of velocities?
This tendency has been shown in recent star-formation simulations \citep{tsukamoto20}.
In this paper, we mainly investigate the red-shifted component to focus on the jet rotation in the high velocity component.
% JMC This paragraph needs further work. 

%%\subsection{Velocity Gradients}
Figure~\ref{f3} shows maps of the mean velocity of the CO ($J$=2--1) emission. 
A clear velocity gradient is found along the jet short-axis in the red-shifted jet (Fig.~\ref{f3}{\it a}).  
The northern part of the red-shifted jet is moving towards the observer after the mean velocity is subtracted and the southern part is moving away from the observer.
The velocity shifts along the minor axis by $25\,\kms$ to $50\,\kms$.
Thus, the red-shifted jet seems to rotate around its long axis.  
%However, except for the rotation,
\S\ref{sec:rotation} discusses alternative interpretations that could create a velocity gradient.
%%s jet precession, twin jets and asymmetrical jet structure \citep{chen16}.

\begin{figure}[!ht]
\centerline{\includegraphics[width=0.95\textwidth]{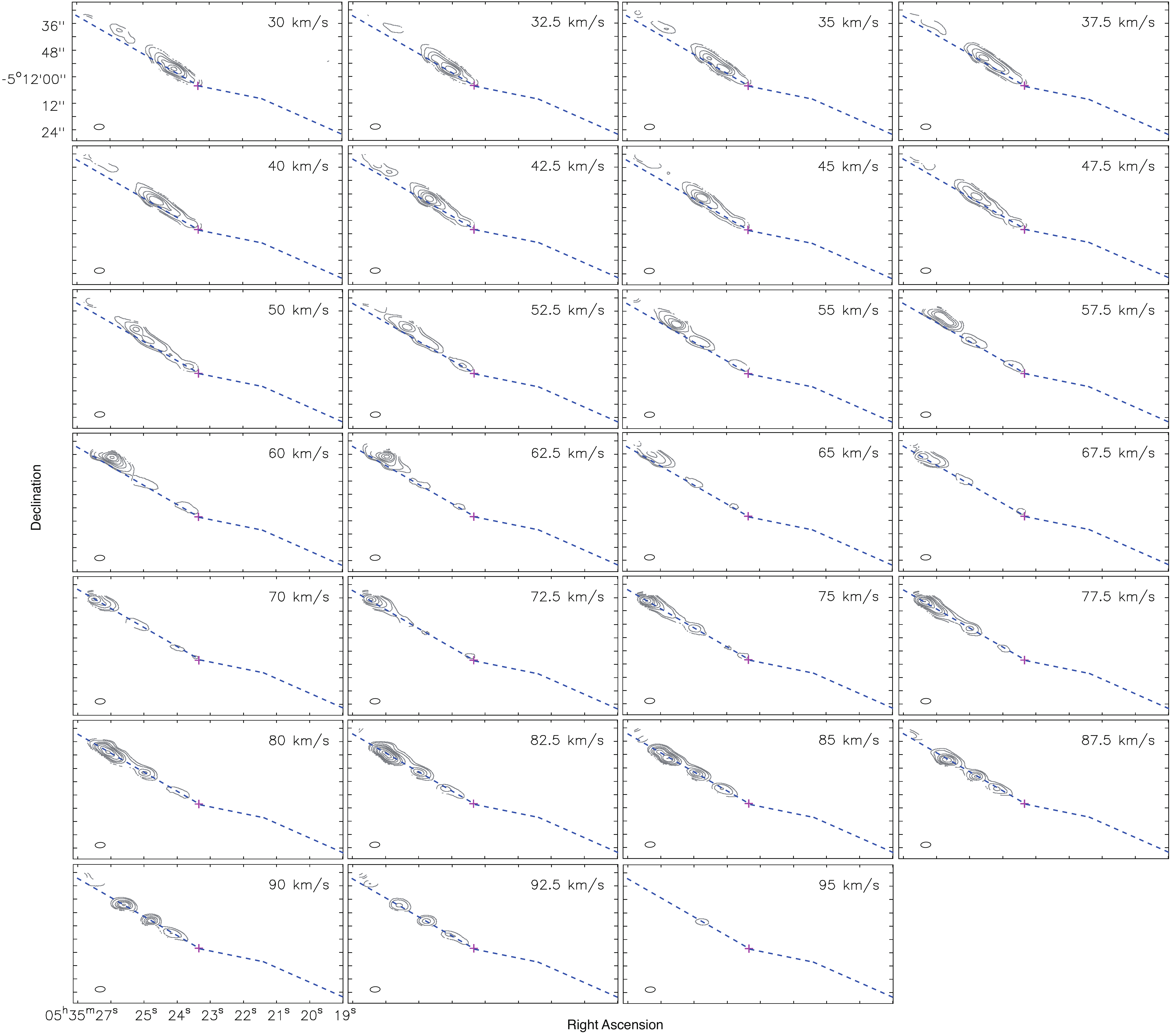}}
\caption 
{CO ($J$=2--1) channel map of the red-shifted flow from FIR 6b.  
The channel maps are presented in the LSR velocity range between $30\,\kms$ and $95\,\kms$, which corresponds to the relative velocity range of $19\,\kms$ to $84\,\kms$ with respect to the systemic velocity ($v_{\rm sys}=11\,\kms$). 
%The channel maps are shown every $2.5\,\kms$ between LSR velocities of $30\,\kms$ and $95\,\kms$ ($v_{\rm LSR}-v_{\rm sys}$ $=$ $19\,\kms$ and $84\,\kms$).
The center value of the LSR velocity is indicated in each panel. 
The plus symbol at the center corresponds to the position of FIR 6b as measured in the 1.3 mm continuum image.
The dashed lines indicate the axes of the red-shifted (northeast side) and blue-shifted  (southeast side) jets.
The jet axis of the red-shifted components is determined from $70.0\,\kms$ channel.
The contour levels are $3\sigma$, $5\sigma$, $10\sigma$, $15\sigma$, $20\sigma$, $30\sigma$, $40\sigma$, $60\sigma$, $80\sigma$, and $100\sigma$ (1$\sigma$ $=$ 30 $\rm{mJy\ beam^{-1}} \cdot \rm{km\ s^{-1}}$).
The black open ellipse in the bottom left corner is the synthesized beam size.
}
\label{f4}
\end{figure}

\begin{figure}[!ht]
\centerline{\includegraphics[width=0.95\textwidth]{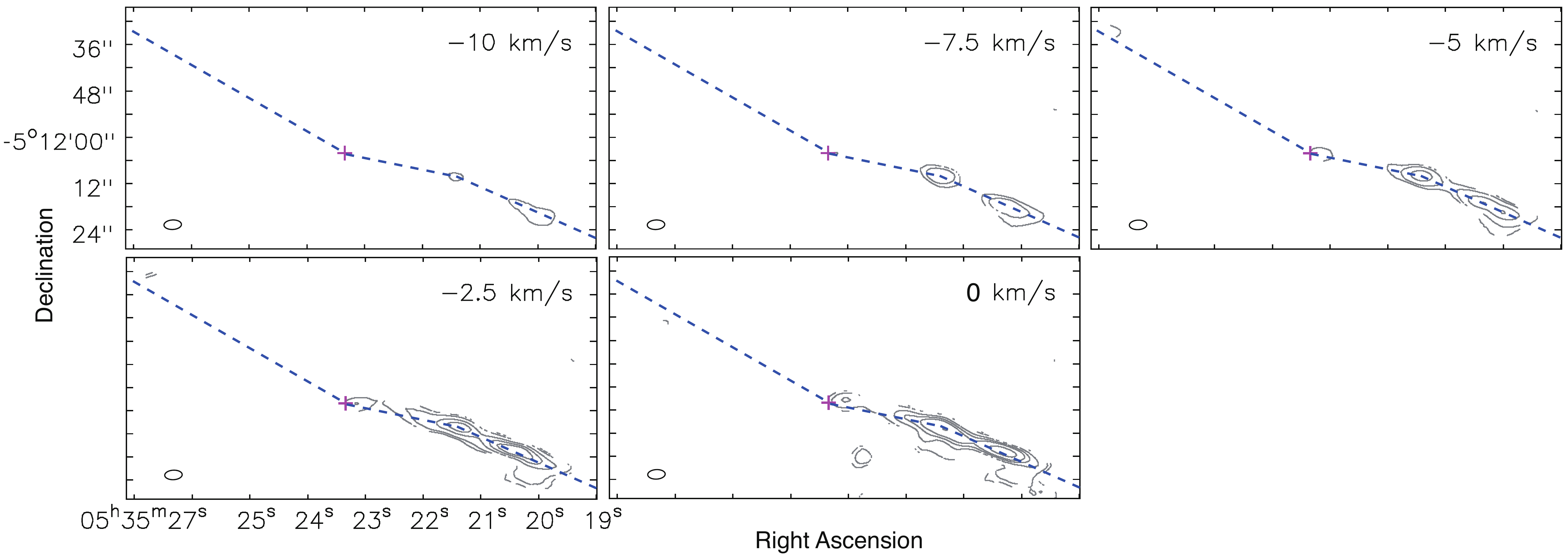}}
\caption{
Same as in Fig.~\ref{f4} but for CO ($J$=2--1) channel map of the blue-shifted flow. 
The channel maps are presented in the LSR velocity range between $-10\,\kms$ and $0\,\kms$, which corresponds to the relative velocity range of $-22\,\kms$ to $-11\,\kms$
% JMC: I think you mean -21 km/s
with respect to the systemic velocity ($v_{\rm sys}=11\,\kms$). 
%As same as in Fig. 5 but for the blue-sifted jet.
%The channel maps are shown every $2.5\,\kms$ between LSR velocities of $-10\,\kms$ and $0\,\kms$ ($v_{\rm LSR}-v_{\rm sys}$ $=$ $-22\,\kms$ and $-11\,\kms$). 
%The channel maps are shown every $2.5\,\kms$ between LSR velocities of -$10\,\kms$ and $0\,\kms$.
%The center value of the LSR velocity is described in each panel. 
%The plus symbol at the center corresponds to the position of protostar FIR 6b measured in the 1.3 mm continuum.
%The broken lines indicate the long axes of the red-shifted (northeast) and blue-shifted (southeast) jets.
%The jet axis of the blue-shifted components is determined on the map of -$5\,\kms$. 
%The contour levels are $3\sigma$, $5\sigma$, $10\sigma$, $15\sigma$, $20\sigma$, $30\sigma$, $40\sigma$, $60\sigma$, $80\sigma$, and $100\sigma$ with 1$\sigma$ =30 $\rm{mJy\ beam^{-1}} \cdot \rm{km\ s^{-1}}$.
%The ellipse in the bottom left corner is the synthesized beam size.
}
\label{f5}
\end{figure}

%%\subsection{Channel map of red-shifted and blue-shifted components.} 
Figures~\ref{f4} and \ref{f5} show the channel maps of the red-shifted (northeast side) and blue-shifted (southwest side) jets.
In Figure~\ref{f4}, the jet axis is defined to pass through the emission peaks of the three knots at a LSR velocity of $70\,\kms$. 
%  JMC: I changed "determined" to "defined".
In the LSR velocity range of $30 - 70\,\kms$, the emission peaks show a deviation from the jet axis in the northeast direction, while in the velocity range of $70 - 95\,\kms$, the deviation from the jet axis is in the southwest direction.
% JMC: Is this true? In the 70-95 km/s range, the emission seems to be coincident with the jet axis.
Although the velocity gradient depends on the distance from the position of FIR 6b, the emission peak is seen on both sides of the jet axis in the channel map.
% JMC: I do not see this.
In addition, we can confirm three knots or relatively strong emission peaks in Figure~\ref{f4}.
% JMC: I suggest to remove the above sentence.
On the other hand, we cannot confirm the emission peak on both sides of the jet axis in the blue-shifted jet (Fig.~\ref{f5}).
% JMC: I suggest to rephrase the above sentence as:
% For the blue-shifted lobe shown in Figure~\ref{f5}, the emission is aligned along a common axis for velocities between $-$10 and 0\,$\kms$. 

\begin{figure}[!ht]
\centerline{\includegraphics[width=1.0\textwidth]{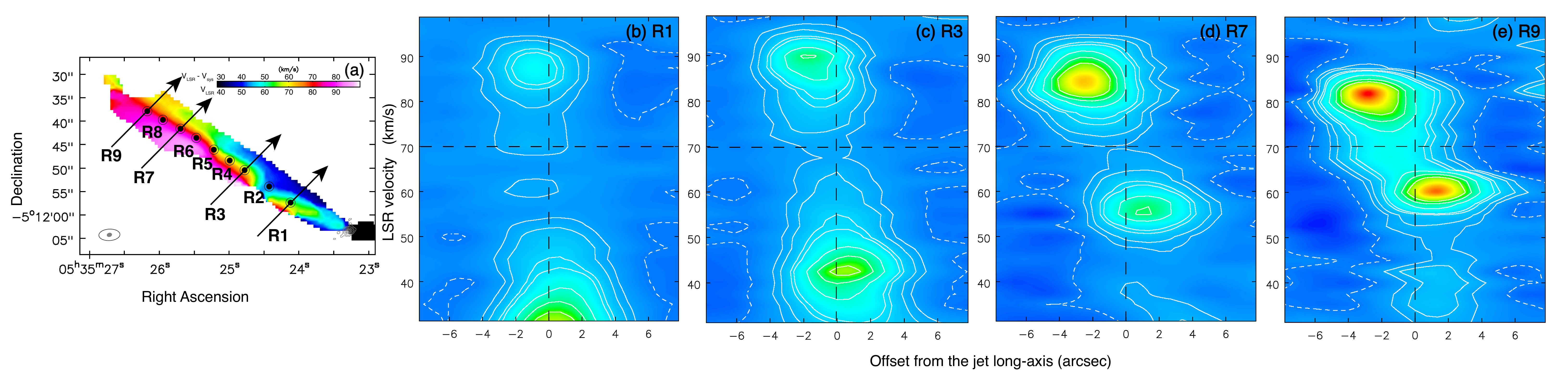}}
%%\centerline{\includegraphics[width=1.0\textwidth]{f3_pb.eps}}
\caption{ 
Position-velocity (PV) diagrams perpendicular to the long axis of the red-shifted jet from FIR 6b.
(a) The mean velocity map of the CO ($J$=2-1) emission. R1--R9 indicate the slices of the position–velocity diagram at 13, 19, 25, 29, 33, 38, 42, 46, and 49 arcsec from the center of the 1.3 mm continuum emission.
Position-velocity diagrams are shown for R1 (panel b), R3 (panel c), R7 (panel d), and R9 (panel e). 
The horizontal dashed line indicates the LSR velocity of $70\,\kms$ (the center speed of jet rotation), while the vertical dashed line corresponds to the center of the jet axis denoted by the filled circles in panel (a).
% JMC: In this figure, you have an open circle surrounding a filled circle. I suggest to remove the open circle and just use a filled circle.
The contour levels are at -5$\sigma$, 5$\sigma$, 10$\sigma$, 20$\sigma$, 25$\sigma$, 50$\sigma$, 75$\sigma$, and 100$\sigma$ (1 $\sigma$ $=$ 10 $\rm{mJy\ beam^{-1}}$).
}
\label{f6}
\end{figure}

%----- [7] PV diagram and quantitative estimate
Figure~\ref{f6} shows  position-velocity (PV) diagrams along the minor axis of the red-shifted component. 
%%The PV diagrams are useful to qualitatively and quantitatively estimate the jet properties.
%Although the velocity gradient along the jet short-axis can be confirmed even in the blue-shifted component (Fig.~2b), it is very clear in the red-shifted component (Fig.~2a). 
%Thus, the northwest jet (or red-shifted jet) is focused to analyze the properties of the jet rotation. 
In Figure~\ref{f6}{\it b}-{\it e}, there are two intensity peaks in the PV diagrams cutting along the short-axis for each red-shifted component at R1, R3, R7, or R9. 
The peaks are located in the second and fourth quadrants independent of the cutting position.
The tendency seen in the PV diagrams agrees well with typical characteristic of rotating cylinder \citep{hirota17}. 
The PV diagrams for FIR 6b (Fig.~\ref{f3}) do not agree with those seen in  numerical simulations of a precessing jet \citep{raga01,masciadri02}.
The two emission peaks would correspond to the wall or edge of the jet where the integrated intensity along the line of sight should be strong. 
%  JMC: Is this for a rotating jet or a precessing jet?
In addition, the directions of velocity gradient are the same in all the cutting positions. 
This is natural if the velocity gradient is attributed to the jet rotation.
% JMC: The sentences starting "The PV diagrams .. " could be moved to Section 4. If you keep them, they need to be revised to make the point clearer.

\section{Jet Rotation and Other Possibilities} \label{sec:rotation}
The velocity gradient detected in the red-shifted component seems to be explained by the jet rotation, as described in \S\ref{sec:results}.
However, other mechanisms can also produce apparent velocity  gradients, such as jet precession, twin jets and asymmetrical jet structure \citep{chen16}. 
In this section, we discuss possible mechanisms that could cause the observed velocity gradient in the FIR 6b jet. 

\begin{figure}[!ht] 
\begin{center}
%\centerline{\includegraphics[width=0.5\textwidth]{f1_pb.eps}}
%%\centerline{\includegraphics[width=0.5\textwidth]{f1_pb.pdf}}
\includegraphics[width=0.9\textwidth]{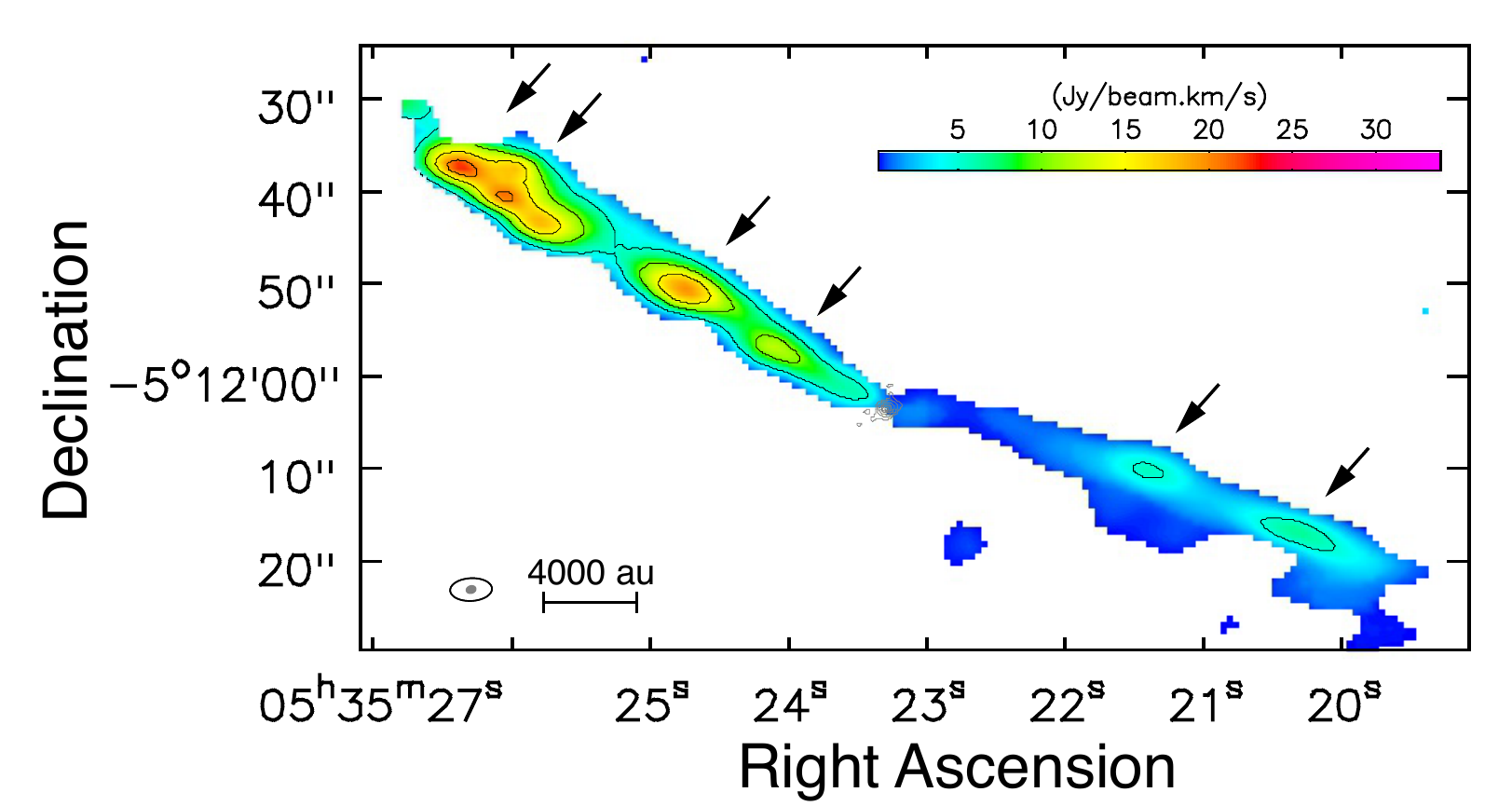}
\caption{ 
CO ($J$=2--1) integrated intensity map of the high velocity  component of the FIR 6b jet (color and black contours).
The emission has been integrated over LSR velocities between 32.5 to 97.5$\,\kms$ (red-shifted component, northeast side) and $-17.5$ to $0\,\kms$ (blue-shifted component, southwest side). 
The  gray contours show the 1.3 mm continuum emission, which peaks peak toward OMC-2/FIR 6b at (R.A., Dec.) = ($05^{h}35^{m}23^{s}.34$, $-05^{\circ}12'03\farcs970$). 
The synthesized beams are shown in the bottom left with a black open ellipse for the CO image and a filled gray ellipse for the 1.3 mm continuum image. 
The contour levels for the CO emission are at 5$\sigma$, 10$\sigma$, 15$\sigma$, and 20$\sigma$ (1$\sigma$ = 1.0 $\rm{Jy\ beam^{-1}} \cdot \rm{km\ s^{-1}}$). Contours for the 1.3 mm continuum emission are at 8$\sigma$, 40$\sigma$, 80$\sigma$, 120$\sigma$, 160$\sigma$, 240$\sigma$, and 320$\sigma$ (1$\sigma$ = 0.2 $\rm{mJy\ beam^{-1}}$).
}
\label{f2}
\end{center}
\end{figure}

\begin{figure}[!ht]
\begin{center}
%%\centerline{\includegraphics[width=0.5\textwidth]{f2_pb.pdf}}
\centerline{\includegraphics[width=0.9\textwidth]{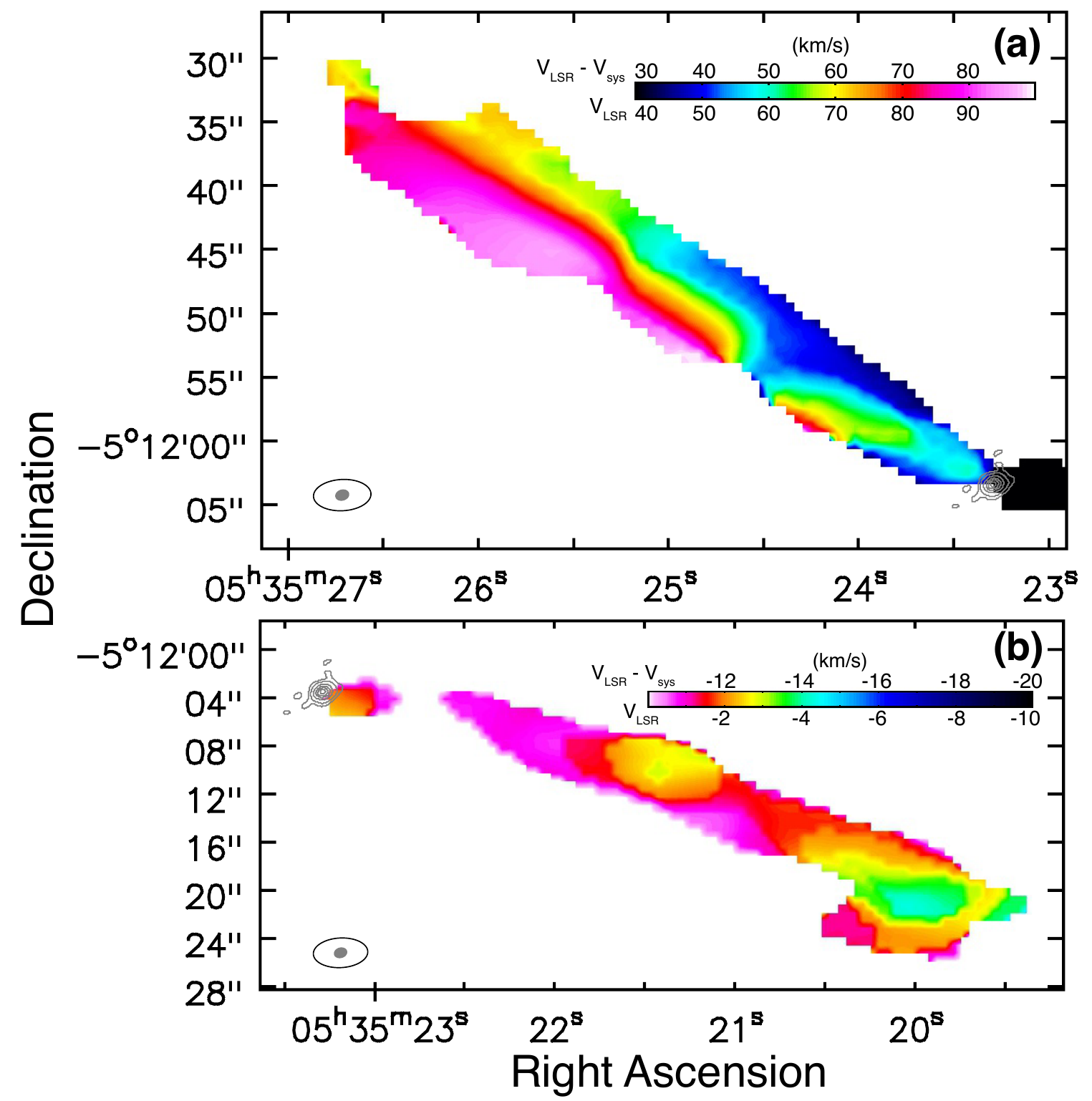}}
\caption{
CO ($J$=2--1) mean velocity maps of the FIR 6b jet.
(a) Velocity map of the northeast side (red-shifted) of the jet, computed for LSR velocities between $32.5~\kms$ to $97.5~\kms$. % excluding the component at the systemic velocity between $v_{\rm LRS} = 7.5\ {\rm and}\ 12.5~\kms$. 
% JMC: I commented this part out since the systemic velocities are not in this range.
(b) Velocity map of southwest side (blue-shifted) of the jet, computed for LSR velocities between of $-17.5~\kms$ to $0~\kms$.
The 1.3 mm continuum emission is also plotted in each panel with gray contours as in Fig.~\ref{f2}.
The ellipses in the bottom left corner are the synthesized beams sized for the CO $J$=2--1  (black) and the 1.3 mm continuum (gray) images.
%, in which the contour levels are set at [8$\sigma$, 40$\sigma$, 80$\sigma$, 120$\sigma$, 160$\sigma$, 240$\sigma$, 320$\sigma$] and 1 $\sigma$ is 0.2 $\rm{mJy\ beam^{-1}}$.
}
\label{f3}
\end{center}
\end{figure}

%----- [4] Interpretation of velocity gradient
When a jet precesses, the velocity gradient tends to appear along the jet long-axis or the jet propagation direction.
On the other hand, the jet velocity is not  significantly changed along the jet short-axis in  precession models \citep{raga01,masciadri02}.
% JMC: I modified the above sentence. Make sure it is ok.
In addition, the precessing jet should have a strong wave-like structure as a whole \citep{masciadri02}, while the observed jet in FIR 6b is straight in both the red-shifted and blue-shifted lobes in the integrated intensity map (Fig.~\ref{f2}).
Moreover, as described in \S\ref{sec:results}, the PV diagrams of the red-shifted component are not in agreement with that derived in  the precessing jet model \citep{raga01,masciadri02}. 
%%The knotty structures within the jets would be attributed to the episodic mass ejection \citep{machida19}.

%----- [5] Interpretation of velocity gradient, binary
If FIR 6b is a binary system, two protostars can drive two jets.
In such a case, the velocity gradient (Fig.~\ref{f3}{\it a}) can be potentially explained by twin jets if the direction of the jets differ slightly  \citep{hara20,saiki20}. 
It seems in fact that the red-shifted jet splits along its long-axis in the velocity map (Fig.~\ref{f3}{\it a}). 
However, we cannot see any signature of the twin jets in the integrated intensity map (Fig.~\ref{f2}).
In addition, there is no evidence for a protobinary system in the 1.3 mm continuum image at 400\,au spatial resolution (Fig.~\ref{f2}) and nor in ALMA 0.87 mm and VLA 9 mm observations at 40\,au resolution \citep{tobin20}.
%In addition, we could not confirm any sign of the existence of a binary system in the dust continuum emission at the root of the jets.
%The spatial resolution estimated from the beam size (1.03 $\times$ 0.78 arcsec) of our continuum data is $\sim400$\,au.
%Furthermore, there is no evidence for a protostellar binary system in other latest observation of FIR 6b with the spatial resolution of $\sim 40$\,au\citep{tobin20}. 
%%Therefore, the  binary system on a scale larger than 40\,au.
%In addition, there is no any observational evidence of binary embedded within FIR 6b.
Thus, it is unlikely that the jet is composed of twin flows. 
%  JMC: I changed "not plausible" to "unlikely".

When the jet has an asymmetric or complex velocity distribution, the velocity gradient would appear in the jet short-axis.
However, the distribution of CO emission is rather smooth (Fig.~\ref{f2}) and the velocity gradient is systematic. 
Thus, the velocity gradients within the FIR 6b jet cannot be explained just by an asymmetric velocity distribution.

%----- [6] Interpretation of velocity gradient, asymmetry
We conclude that rotation is the most plausible interpretation of the velocity gradient observed in the FIR 6b jet (Fig.~\ref{f3}{\it a}). 
%  It is difficult to explain the velocity gradient with other models except for rotation. 
% JMC: I removed this sentence since was repetative.
In addition, there are similarities between numerical simulations of protostellar jets and the observations. 
A winding configuration of velocity gradient seen in the velocity map can be seen in  simulations of rotating jets \citep{staff15}.
%  JMC: I did not  understand what was meant by "a winding configuration" since the term "winding" is not used elsewhere in the text.
Further, the knots along the jet seen in the integrated intensity map (Fig.~\ref{f2}) are also produced in simulations and can be explained by time-variable mass ejection \citep{machida19}.

\section{Derivation and Comparison of Jet Parameters} 
\label{sec:parameters}
In this section, we derive the properties of the red-shifted jet. 
Using the position-velocity diagram (Fig.~\ref{f6}), the jet rotation velocity $v_{\phi}$ and the distance from the jet axis (or radius in the cylindrical coordinate) $r_{\rm rot}$ are estimated as 
\begin{equation}
\label{eq:vel}
v_{\phi} = \frac{1}{{\rm sin}\ i}\frac{v_{\rm blue}-v_{\rm red}}{2},
\end{equation}
%%\begin{equation}
%%v_{\rm rot} = \frac{1}{{\rm sin}\ i}\frac{v_{\rm shift}}{2},
%%\end{equation}
and 
\begin{equation}
r_{\rm rot} = \frac{l_{\rm shift}}{2},
\end{equation}
respectively \citep{chen16}, where $v_{\rm blue}$ and $v_{\rm red}$ are the velocity at emission peak of the red-shifted and blue-shifted components in each PV diagram (Fig.~\ref{f6}).
%%The jet rotation axis was determined using the PV diagrams, in which the mid point between the two emission peaks is adopted as the center of the jet axis.
The jet rotation axis was determined using the PV diagrams, in which the mid point between the two emission peaks is adopted to define  the jet axis.
% JMC: Check if the above sentence is correct.
The jet width $l_{\rm shift}$ was calculated as the distance between the two peaks in the PV diagrams.
%%the velocity gradient $v_{\rm shift}$ was estimated as the velocity difference between two peaks and

\citet{furlan16} estimated a disk inclination angle of $i=80^{\circ}$ with respect to the line of sight \citep{furlan16}. Using our 1.3 mm continuum data, we fitted a two-dimensional Gaussian to derive a deconvolved disk size of $L_{\rm maj} \times L_{\rm min} = 1\farcs1 \times 0\farcs82$ ($416  \times 312$ au), which implies an inclination angle of $i = {\rm arccos}(L_{\rm min}/L_{\rm maj}) = 42^{\circ}$.
%%(P.A. $= 109^{\circ}$) with 1.3 mm continuum emission 
\citet{tobin20} using higher angular resolution observations to derive a disk inclination angle of $i=55^{\circ}$.
% Therefore, the jet inclination angles were estimated to $40^{\circ}$, $60^{\circ}$, and $80^{\circ}$ in this and other observations.
Recent theoretical study showed that the inclination angle of the disk (or normal direction of the disk) may change depending on the spatial scale \citep{machida20}.
Thus, it is difficult to accurately determine the inclination angle of the disk and jet from observations. 
%  JMC: Given you do not  quote any uncertainties in the disk inclination angle, I just to remove the above two sentences.
Here, we adopt the jet inclination angle $i=80^{\circ}$, which gives the minimum rotation velocity and specific angular momentum (see eq.~[\ref{eq:vel}]). 
We discuss the dependence of the inclination angle on the jet properties in \S\ref{sec:inclination}.

%%As a representative number, we using $80^{\circ}$, which is the minimum value of the results of $j$ and $v_\phi$.

The measured rotation velocities are in the range of $v_{\phi} = 11.2 - 28.4\,\kms $, while the jet radius (or the distance from the jet axis) is in the range of $r_{\rm rot}\sim220-800$\,au (Table~\ref{tab1}).
The rotation velocities estimated for FIR 6b are the highest among those reported in other sources even after correcting for the jet inclination angle: $v_{\phi}=6 - 15\,\kms$ for DG Tau \citep{bacciotti02},  $\sim 2\,\kms$ for CB 26 \citep{Launhardt09}, $\sim 5 - 14\,\kms$ for SVS 13A \citep{chen16}, and $\sim 10\,\kms$ for Orion Source I \citep{hirota17}. 

Given the inferred jet velocity and radius, the specific angular momentum of the jet at each position can be estimated as  $j =v_{\rm rot} \times r_{\rm rot}$.
The specific angular momentum is in the range of $9.2\times10^{21}\,\jj \le j \le 1.9\times10^{22}\,\jj$ (Table 1).
The maximum specific angular momentum estimated in this study is one or two orders of magnitude larger than those reported in other sources.
The specific angular momenta calculated in previous studies are $j\sim 3.5 \times 10^{20}\,\jj$ for DG Tau \citep{bacciotti02}, $\sim 1.5 \times 10^{20}\,\jj$ for CB 26 \citep{Launhardt09} and $\sim 7.5 \times 10^{20}\,\jj$ for Orion Source I \citep{hirota17}. 
Among the jets where the rotation speed was measured, only the rotating jet bullet SVS 13A has a comparable but slightly smaller value of the specific angular momentum  ($\sim 9.8 \times 10^{21}\jj$) \citep{chen16}.
%%Chen et al. 2016 imply that rotating jets from episodic outbursts in the protostellar phase can more efficiently remove the angular momentum from the disk. 
 
It is crucial to explore the launching radius $r_0$ of the jet in order to clarify the jet driving mechanism.
Based on Bernoulli's theorem and conservation of angular momentum, we can estimate the jet launching radius \citep{mestel68}. Using the same method, the launching radius ($r_0$) has been estimated as 10-100\,au for low-velocity outlows \citep{hirota17,alves17} and 0.05-7\,au for high-velocity jets \citep{lee17, chen16} in various objects.
%recent ALMA observations estimated the launching radius $r_0$ of both low-velocity outflow and high-velocity jet for different objects; $r_0 \sim 10-100$\,au  for the low-velocity outflows \citep{hirota17,alves17} and $r_0 \sim0.05-7$\,au for the high-velocity jets \citep{lee17, chen16}. 
For FIR 6b, the launching radius is estimated using the toroidal $v_\phi$ and poloidal $v_{\rm pol}$ velocities and the jet radius of each position with the analytical solution \citep{anderson03}
\begin{eqnarray}
r_{0} &=& 0.7\ {\rm au} \left(\frac{r_{\rm rot}}{10\ {\rm au}}\right)^{2/3} \left(\frac{v_{\phi}}{10\ {\rm km\ s^{-1}}}\right)^{2/3}  \left(\frac{v_{\rm pol}}{100\ {\rm km\ s^{-1}}}\right)^{-4/3} \left(\frac{M_{\rm star}}{1\ M_\odot}\right)^{1/3},  
\label{eq:r0}
\end{eqnarray}
where the poloidal velocity components of the jet  are derived from 
\begin{equation}
v_{\rm pol} = \frac{1}{{\rm cos}\ i}\frac{v_{\rm blue}+v_{\rm red}}{2}.
\end{equation}
The mass of the central protostar is  assumed to be $M_{\rm star}=$ 1 $M_\odot$.
Note that the dependence of the estimated launching radius on the protostellar mass is weak (eq.~[\ref{eq:r0}]).
The rotation velocity and radius estimated in the PV diagrams imply a jet launching radius in the range of $r_0 = 2.18-2.96$\,au (Table~\ref{tab1}). 
When $M_{\rm star}=$ 0.1 $M_\odot$, then $r_0 = 1.01-1.38$\,au.
Thus, the jet is expected to be driven by the intermediate region of the circumstellar disk where magnetic field is expected to be coupled with neutral gas \citep{machida19}, not by the region very close to the protostar. 
Therefore, these observations are most consistent with the disk wind mechanism \citep{blandford82,tomisaka02}, where the jet is driven from a wide range of radii. The observations are inconsistent with the X-wind and entrainment mechanisms \citep{shu94,Arce07}, in which the jet is driven from a region very close to the protostar ($\lesssim 0.01$\,au).
% JMC: Check that the last couple of sentences are OK.
High-velocity jets with a small amount of angular momentum are expected in the X-wind and entrainment scenarios, because the jets appear in the region near the protostar where the angular momentum is not large.

The very large specific angular momentum of the FIR 6b jet cannot be easily explained.
For example, the centrifugal radius is estimated as $r_{\rm cent}=j^2/(GM_{\rm star})$, inside which the Keplerian rotation is assumed.
With $j=10^{22}\,\jj$ and $M_{\rm star}=1\msun$, the centrifugal radius becomes $r_{\rm cent}=4.9\times10^4$\,au. 
The centrifugal radius estimated here is much larger than the disk-like structure observed around FIR 6b, and is not realistic. 
Thus, the jet needs to obtain a very large angular momentum through some other mechanism. The magnetic effect can provide a reasonable solution to the large specific angular momentum and the origin of the super-rotation of the jet.
% JMC: I suggest to remove the word "Only" since you have not  really demosntrated this is the *only* solution.

To investigate the effect of magnetic field on the angular momentum transfer, the Alfv\'en radius $r_{\rm A}$ \citep{anderson03} at each position is estimated using
\begin{equation}
    r_{\rm A} = \sqrt{\frac{r_{\rm rot} v_\phi}{\Omega_0}},
\end{equation}
where $\Omega_0 = \sqrt{GM_{\rm star}/r_0^3} $ is the Keplerian angular velocity at the jet launching radius.
The Alfv\'en radius is in the range of $r_{\rm A}=25.7-46.5$\,au (Table~\ref{tab1}).
The ratio of the Alfv\'en radius to the jet launching radius ($\lambda$) is as large as $\lambda \equiv {r_{\rm A}/r_0}=11.8-15.7$.
Thus, the long magnetic lever arm efficiently transports the angular momentum from the circumstellar region for the super-rotating jet case \citep{pudritz19}. 

The magnetic pressure dominates the ram pressure within the Alfv\'en radius. 
Therefore, the large $\lambda$ means that the magnetic field plays a dominant role near the protostar. 
The large lever arm is realized when the magnetic field threading the disk is strong and the mass loaded onto the wind (or the magnetic field line) is very small under which the jet  can be accelerated to a large cylindrical radius before the magnetic field line is bent back by the plasma inertia. 
We expect such an environment is realized in a late main accretion phase during which the plasma beta (the ratio of  thermal to magnetic pressure)  above and below the disk is expected to be very low with a low-density and a strong magnetic field, and the wind mass is also very small \citep{machida13}.

%%have a low density and a strong magnetic field, and the wind mass is also very small \citep{machida13}.
% JMC: What is meany by "plasma beta">
%  JMC: suggest "expected to have a low density and a strong magnetic field..."

%%The large lever arm is expected to be realized in a late main accretion phase during which the density of the infalling envelope should be low.
%%The low density causes a small ram pressure, while the magnetic field should be strong near the protostar. 
%%As a result, the magnetic-dominated region expands as the infalling envelope dissipates or the density of the infalling envelope lowers. 

\begin{figure}[!ht]
\centerline{\includegraphics[width=0.6\textwidth]{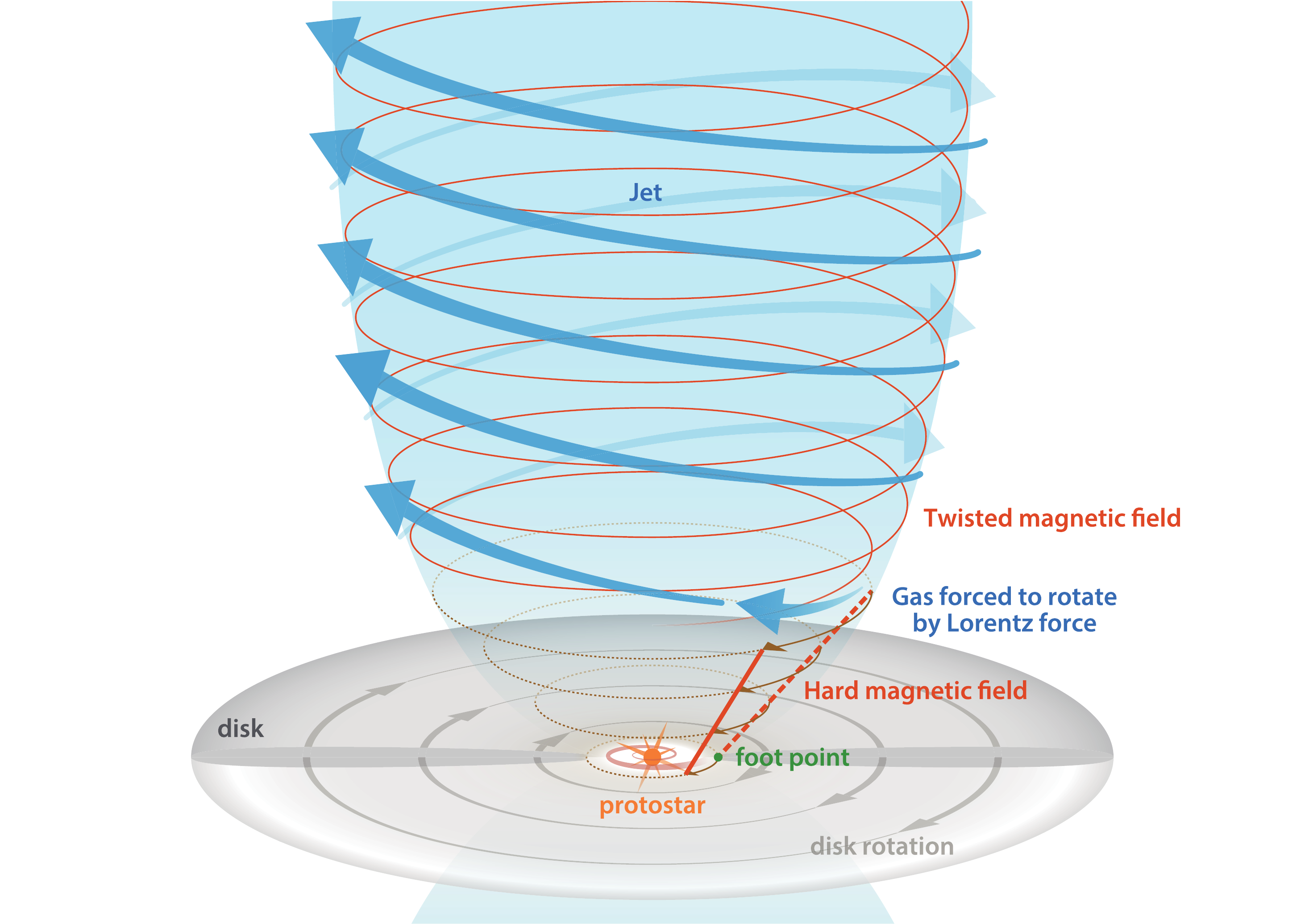}}
\caption{ 
Schematic view of a super-rotating jet. 
The magnetic field line inside the Alfv\'en radius (thick dotted and solid red lines) rotates rigidly with the Keplerian velocity at its foot point and accelerates the gas elements located at the tip. 
The magnetic field line (red curve) is strongly twisted within the jet.
The angular velocity of the jet is the same as that of the foot point  and thus the super-rotation (blue arrows) is realized. 
}
\label{f7}
\end{figure}

Since the jet launching region is in the range of $2.18-2.96$\,au (Table~\ref{tab1}), the foot points of the magnetic field lines connecting to the jet are also distributed in the same range of the circumstellar disk. 
% JMC: You need to define "foot points".
Thus, wirelike strong (or hard) magnetic field lines originating in the circumstellar disk near the protostar ($2.18-2.96$\,au) swing the fluid elements located very far from the foot points of the magnetic field lines ($25.7-46.5$\,au). 
Since the fluid elements are frozen to the magnetic field lines, they are forced to corotate with the Keplerian velocity at foot points of the magnetic field lines, as seen in the schematic view of Figure~\ref{f7}.  
As a result, the fluid elements receive the angular momentum and are expelled from the circumstellar region by magnetic effect \citep{pudritz86}.
On the other hand, a parcel of gas in the circumstellar disk near the protostar loses the angular momentum and falls onto the protostar \citep{pudritz19}. 
Therefore, the excess angular momentum is ejected from the circumstellar disk by the rotating jet, and the gas whose angular momentum has been removed by the jet falls onto the protostar and promotes the protostellar growth.  

\section{Discussion}
\subsection{Effect of Jet Inclination Angle}
\label{sec:inclination}
The inclination angle of the jet is adopted as $i=80^{\rm \circ}$ when estimating the properties of the jet.
However, as described in \S\ref{sec:parameters},  the inclination angle of the disk is inferred to be $i=42^{\rm \circ}$ from our observations of the 1.3 mm continuum and $i=55^{\rm \circ}$ from published, higher angular resolution continuum observations \citep{tobin20}.

Since the jet physical quantities depend on the assumed inclination angle, Table~\ref{tab2} lists the jet properties for $i=40^{\rm \circ}$ and $60^{\rm \circ}$.
For these two cases, the specific angular momentum exceeds $j=10^{22}\ \jj$, the maximum rotation velocity  exceeds 40\,$\kms$, and the
jet launching radius ranges between 4 and 13\,au, which is considerable further from the star when assuming $i=80^{\rm \circ}$.
Therefore the adopted inclination angle does not significantly affect the conclusions.
Thus, the distant launching radius and the rapid jet rotation inferred for the FIR 6b jet imply that the X-wind and entrainment scenarios are not the primary driving mechanisms.
% JMC: Double-check to make sure you are ok with this sentence.
It should be noted that although the CO outflow discussed in this study supports the disk wind scenario, an X-wind component could still present.
If future observations identify a faster component with a small amount of angular momentum, it might correspond to the jet in the X-wind model. 
The disk wind and X-wind can coexist and contribute to the angular momentum transfer at different radii in the disk. 
% JMC: Double check that you are happy with the last two sentences.

\subsection{Low-velocity Component} 
% As seen in Figure~\ref{f2}, the rotation motion is confirmed only in the high velocity component of $20-85\,\kms$ with respect to the systemic velocity.
%In contrast to the high-velocity components, the velocity gradient along the short axis of the jet cannot be identified in the low-velocity components (Fig.~\ref{f1}).
% Thus, there is no clear sign of the rotation 
%As seen in Figure~\ref{f2}, a clear velocity gradient is only observed in the red-shifted, high-velocity component. 
No clear velocity gradient is detected in either of the low-velocity components (Fig.~\ref{f1}).
The outflow (or jet) velocity should be proportional to the Keplerian velocity at the outflow driving radius.
Thus, it is natural that rapid rotation is observed only in the high-velocity component, in which the base  of the high-velocity outflow (or jet) is located near the protostar where the Keplerian rotation is high. 
The lack of a detectable velocity gradient in the low-velocity components may be attributed to a driving radius located relatively far from the central protostar where the Keplerian rotation velocity is small. 
Further high resolution observations are needed  to confidently understand the driving mechanism of the jets. 

\section{Summary}
We present high resolution ALMA observations of CO ($J$=2--1) and the 1.3 mm continuum of the protostellar jet from the protostar FIR 6b in OMC-2. A clear velocity gradient is found along the short axis of the high-velocity, red-shifted component of the jet. We attribute the velocity gradient to rotation.
Using PV diagrams and an analytical model, the jet properties are inferred. 
The rotation velocity ($>20\,\kms$) and specific angular momentum ($>10^{22}\,\jj$) are extraordinarily large compared to previously observed jets. We suggest that the super-rotation of the jet is caused by the efficient transfer of angular momentum from the disk to the jet by magnetic fields. In addition, the jet launching points are inferred to be a radius of $2.18-2.96$\,au.
These findings are consistent with the disk wind hypothesis as the primary jet driving mechanism for FIR 6b. 
The study could unveil the final phase of the protostellar evolution.
%  JMC: Suggest to remove the last sentence since this point  is not discussed in the paper.

%%First of all, we need to determine the jet inclination angle to investigate the jet parameters. 
%%We adopted the jet inclination angle $i=80^{\rm \circ}$ with respect to the line of sight \citep{furlan16}.

%\begin{addendum}
%----- Acknowledgements
\acknowledgments
We thank the referee for very useful comments and suggestions on this paper. 
This work was financially supported by the Grants-in-Aid for JSPS Fellows (YM).
This paper makes use of the following ALMA data: ADS/JAO.ALMA\#2017.1.01353.S. ALMA is a partnership of ESO (representing its member states), NSF (USA) and NINS (Japan), together with NRC (Canada), MOST and ASIAA (Taiwan), and KASI (Republic of Korea), in cooperation with the Republic of Chile. The Joint ALMA Observatory is operated by ESO, AUI/NRAO and NAOJ. The present study was supported by JSPS KAKENHI grants (JP21K03617, JP21H00046: MNM)

\bibliography{bib}{}

\clearpage
%----- Table 1, Jet rotation  parameters
\begin{table}[t]
\begin{center}
\caption{Physical properties of the jet for $i = 80^{\circ}$}
\label{tab1}
%%\scalebox{0.85}{
\begin{tabular}{ccccccccc} \hline
Position & $j$  & $r_0$ &  $r_{\rm rot}$  & $v_{\phi}$ & $v_{\rm p}$ & $r_{\rm A}$ \\
%& ($i$ = $80^{\circ}$) & ($i$ = $80^{\circ}$) & & ($i$ = $80^{\circ}$) & ($i$ = $80^{\circ}$) & ($i$ = $80^{\circ}$)\\
& [$\rm cm^2\ s^{-1}$] &  [au]  & [au] & [$\rm km\ s^{-1}$] & [$\rm km\ s^{-1}$] & [au] \\ \hline
R 1& 9.2$\times  10^{21}$ & 2.18 & 216 & 28.4 & 344 & 25.7 \\
R 2 & 1.6$\times  10^{22}$ & 2.79 &440 & 23.9 & 364 & 40.5\\
R 3 & 1.9$\times  10^{22}$ & 2.96 &529 & 23.9 & 381 & 46.4\\
R 4 & 1.4$\times  10^{22}$ & 2.53  &480 & 19.3 & 367 & 35.4\\
R 5 & 1.7$\times  10^{22}$ & 2.78 &700 & 16.2 & 379 & 41.9\\
R 6 & 1.5$\times  10^{22}$ & 2.35  &549 & 18.8 & 410 & 35.3\\
R 7 & 1.7$\times  10^{22}$ & 2.60  &760 & 15.2 & 402 & 40.3 \\
R 8 & 1.5$\times  10^{22}$ & 2.32  &800 & 12.2 & 402 & 34.0\\
R 9 & 1.4$\times  10^{22}$ & 2.16 &804 & 11.2 & 407 & 31.0\\ \hline 
\end{tabular}
%%}
\end{center}
\end{table}

\begin{table}[t]
\setlength{\tabcolsep}{2.5pt}
\begin{center}
\caption{Physical properties of the jet for $i=40$ and $60^\circ$}
% JMC: I changeed the title to be consistent with Table 1.
\label{tab2}
\scalebox{0.85}{
\begin{tabular}{c|ccccc|ccccc}\hline 
& &  & $i$ = $40^{\circ}$ &  & &  &  & $i$ = $60^{\circ}$ & & \\
Position & $j$ & $r_0$ & $v_{\phi}$ & $v_{\rm p}$ & $r_{\rm A}$ & $j$ & $r_0$ & $v_{\phi}$ & $v_{\rm p}$ & $r_{\rm A}$\\
& [$\rm cm^2\ s^{-1}$] &  [au]  & [$\rm km\ s^{-1}$] & [$\rm km\ s^{-1}$] & [au] & [$\rm cm^2\ s^{-1}$] &  [au]  & [$\rm km\ s^{-1}$] & [$\rm km\ s^{-1}$] & [au]\\ \hline
% & [$\rm cm^2\, s^{-1}$] &  [au] & [$\rm km\,s^{-1}$] & [au] & [$\rm cm^2\, s^{-1}$] & [au] & [$\rm km\, s^{-1}$] & [au]\\ \hline
R1 & 1.4$\times 10^{22}$ & 20.1 & 43.6 & 78.3 & 169 & 1.0$\times 10^{22}$ &9.3  & 32.3& 120& 81.5 \\
R2 & 2.4$\times 10^{22}$ & 26.6 & 36.6 & 82.9 & 272 &1.8$\times 10^{22}$ & 12.3 & 27.1& 127 & 120\\
R3 & 2.9$\times 10^{22}$ & 28.3 & 36.6 & 86.8 & 313 &2.2$\times 10^{22}$ & 13.1 & 27.1 & 133 & 152 \\
R4 & 2.1$\times 10^{22}$ & 24.2 & 29.6 & 83.5 & 238 &1.6$\times 10^{22}$ & 11.2 & 21.9 & 128 & 115\\
R5 & 2.6$\times 10^{22}$ & 26.7 & 24.9 & 86.1 & 284 &1.9$\times 10^{22}$ & 12.4 & 18.5 & 132 & 138\\
R6 & 2.4$\times 10^{22}$ & 22.5 & 28.8 & 93.3 & 238 &1.8$\times 10^{22}$ & 10.4 & 21.3 & 143 & 115\\
R7 & 2.7$\times 10^{22}$ & 24.9 & 23.3 & 91.4 & 272 &2.0$\times 10^{22}$ & 11.6 & 17.3 & 140 & 132\\
R8 & 2.2$\times 10^{22}$ & 22.2 & 18.7 & 91.4 & 229 &1.7$\times 10^{22}$ &10.3 & 13.9 & 140 & 111\\
R9 & 2.1$\times 10^{22}$ & 20.6 & 17.1 & 92.7 & 208 &1.5$\times 10^{22}$ & 9.6 & 12.7 & 142 & 101\\ \hline
\end{tabular}
}
\end{center}
\end{table}

\clearpage
%----- Fig. 1

%----- Fig. 2

%----- Fig. 3

%----- Fig. 4

%----- Fig. 5

%----- Fig. 6

%----- Fig. 7

\end{document}